\documentclass[12pt]{article}

\usepackage{amsmath}
\usepackage{amsfonts}
\usepackage{latexsym}
\usepackage{a4wide}
\usepackage{euscript}
\usepackage{exscale}

\newcommand{\M}{\mbox{${\cal M}$}}
\newcommand{\N}{\mbox{${\cal N}$}}
\newcommand{\D}{\mbox{${\cal D}$}}
\newcommand{\E}{\mbox{${\cal E}$}}
\renewcommand{\O}{\mbox{${\cal O}$}}
\newcommand{\A}{\mbox{${\cal A}$}}
\newcommand{\RW}{Robertson-Walker spacetime }
\newcommand{\RWs}{Robertson-Walker spacetimes}
\newcommand{\EuS}{\mbox{${\EuScript S}_\varepsilon$}}
\newcommand{\skalar}[2]{\mbox{$\langle #1 | #2 \rangle$}}
\newcommand{\Cnull}[1]{\mbox{$C_{0,\mathbb R}^\infty(#1)$}}
\newcommand{\ol}{\overline}

\newtheorem{theo}{Theorem}
\newtheorem{defi}{Definition}

\begin{document}

\title{Correlations of Quantum Fields on\\ Robertson-Walker Spacetimes}
\author{Mathias Trucks\thanks{\hspace{0.1cm}
    e-mail: mathias@itpos22.physik.tu-berlin.de}\\
       \small Institut f\"ur Theoretische Physik,\\
       \small Technische Universit\"at Berlin\\
       \small Hardenbergstra{\ss}e 36, 10623 Berlin, Germany\\[0.3cm]}
\maketitle

\begin{abstract}
  It is a well known fact that quantum fields on Minkowski spacetime
  are correlated for each pair of spacetime regions. In \RWs{} there are
  spacelike separated regions with disjoint past horizons but
  the absence of correlations in that case was never proved.
  We derive in this paper formulae for correlations of quantum fields
  on \RWs. Such correlations could have reasonably influenced the
  formation of structure in the early universe. We use methods of
  algebraic and constructive quantum field theory.\\[0.5cm]
\end{abstract}

\section{Introduction}

In this paper we derive formulae for correlations of quantum fields on
\RWs. We follow an idea of R.~M.~Wald \cite{Wald2} who argued that
such correlations could be of importance for formation of structure in
the early universe.

Considering local properties of quantum fields on Minkowski
spacetime, it is a well known fact that correlations are always
present, even for spacelike separated regions. On \RWs{} there are in
contrast to Minkowski spacetime spatially separated regions whose
past horizons do not intersect. It is usually taken for granted that
events occurring in such regions are statistically
independent. However, up to
our knowledge the absence of correlations was never explicitly
proved.

The question whether such correlations do exist may have important
relevance in cosmology.
A typical example for the application of the argument that events in
regions with disjoint past horizons are totally independent, is
the formation of topological defects in the early universe, such as
magnetic monopoles, cosmic strings and domain walls. Such defects are
believed to occur if the universe undergoes a phase transition.

We work in the framework of algebraic quantum field theory and use
methods of constructive quantum field theory for the derivation of
the formulae. To begin with, we consider a neutral free boson field on a
\RW and hope to extend this method to more complicated quantum field
theories.

The point of the paper is the following. Usually the presence of
field correlations in two spatially separated regions is expressed by
the formula
\[
   <\phi(f)\phi(g)> \;\; \neq \;\; <\phi(f)><\phi(g)> \, ,
\]
where $\phi$ is the field operator and $f,g$ are two test functions
with spacelike separated supports. The
inequality just tells us that correlations are
present.

Complete knowledge
about the correlations is obtained if the joint probability of two
events to occur in the two spacelike separated regions, respectively,
is compared with the product of the probabilities for their respective
occurences. Let ${\cal B}_1,{\cal B}_2$ be any pair of Borel sets on
the real line and $\E_f,\E_g$ be the spectral measures belonging to
the operators $\phi(f),\phi(g)$, respectively. Then the measures
$<\E_f\E_g>$ and $<\E_f><\E_g>$ have to be compared. The events ${\cal
  B}_1$ and ${\cal B}_2$ are correlated if
\[
   <\E_f({\cal B}_1) \E_g({\cal B}_2)> \;\; \neq \;\; <\E_f({\cal
     B}_1)><\E_g({\cal B}_2)>\, ,
\]
otherwise they are uncorrelated. This means we are interested in a
spectral decomposition of the field operator.

A drawback is the fact that the formulae can not be solved
analytically for fields located in spacelike separated regions. They
can only be solved numerically and related to this we can not analyse
the influence of the past horizon.

At present we are not able to compute such correlations for an interacting
theory but take this work as a first step. Of course it would be more
interesting to derive formulae for a $\phi^4$-theory since the formation
of topological defects is expected by such theories. However, we wish to
point out that correlations in a noninteracting theory are most
unexpected because they can not be considered as a result of a nontrivial
interaction. Correlations in a noninteracting theory would strongly
suggest the occurrence of stronger correlations in an interacting theory.

The paper is organized at follows. In section 2 we discuss the
classical solutions of the Klein-Gordon equation on a Robertson-Walker
spacetime. In the next section we construct the Weyl algebra
corresponding to the spaces of classical solutions obtained in section
2. In section 4 a distinguished class of states, the adiabatic vacuum
states, on these Weyl algebras are discussed. The computation of
correlations will be done in these states which are known to be {\em
  good} states for these spacetimes. Since we are interested in a
spectral decomposition of the field operator we introduce Gaussian
measures on a functional Hilbert space in section 5 and decompose the
field operator in section 6 with the help of these spaces. Also in
section 6 we summarize the formulae for the correlations.

\section{The spaces of classical solutions}
\label{section_classical}

We discuss the classical solutions of the Klein-Gordon equation
\[
   (\Box_g - m^2)\varphi = 0 \; ,
\]
on a Lorentz manifold $(\M,g)$, where $g$ is a
Robertson-Walker metric
\[
        g = -dt^2 + R(t)^2[d\theta_1^2 + \Sigma^2_\varepsilon(d\theta_2^2
             + \sin^2\theta_2 d\phi^2)] \, , \quad R(t) > 0.
\]
These are homogeneous and isotropic spaces, topologically of the form
$\M = {\mathbb R} \times \EuS$, where $\varepsilon=1,0,-1$ corresponds
to the spherical, the flat and the hyperbolic case, respectively.We
think of the spatial parts as embedded in ${\mathbb R}^4$ by
\begin{align*}
  {\EuScript S}_1 &= \{ x \in {\mathbb R}^4 | \; \sum_{i=0}^3
  x_i^2  = 0\} \, , & \Sigma_1 & = \sin\theta_1 \, , \\
  {\EuScript S}_0 &= \{ x \in {\mathbb R}^4 | \; x_0 = 0 \} \; ,
  &  \Sigma_0 & = \theta_1 \, ,\\
  {\EuScript S}_{-1} &= \{ x \in {\mathbb R}^4 | \; x_0^2 -
  \sum_{i=1}^3 x_i^2 = 1, \; x_0 > 0 \} \, , & \Sigma_{-1} & =
  \sinh\theta_1 \, .
\end{align*}
The Klein-Gordon operator on these spaces takes the form
\[
    \Box_g - m^2 =
        -\frac{\partial^2}{\partial t^2}
        -3 \frac{\dot{R}(t)}{R(t)}
        \frac{\partial}{\partial t}+ \frac{1}{R^2(t)}
        \Delta_\varepsilon - m^2           \, ,
\]
where $ \Delta_{\varepsilon}$ means the Laplace operator on the
respective spatial parts.

\subsection{The spherical case}
The eigenfunctions of the Laplace operator on the 3-sphere
$\EuScript{S}_1$ are functions
\[
   Y_{\vec{k}}(\vec{x}) = e^{\pm i k_2 \phi} (\sin\theta_1)^{k_1} \,
        C_{k_0\!-k_1}^{k_1\!+1}(\cos\theta_1) (\sin\theta_2)^{k_2} \,
        C_{k_1\!-k_2}^{k_2\!+1/2}(\cos\theta_2) \, ,
\]
$\vec{x} = (\theta_1, \theta_2, \phi)$, $\vec{k} = (k_0,k_1,k_2)$,
$k_0 \in {\mathbb N}_0$, $0<k_1<k_0$, $-k_1<k_2<k_1$ and
$C_n^l$ are Gegenbauer polynomials (see e.g.~\cite[Ch.~11.2]{EMOT2}).
These eigenvectors fulfill the equation
\[
    \Delta_1 Y_{\vec{k}}(\vec{x}) =
       -k_0(k_0 +2)Y_{\vec{k}}(\vec{x}) \, ,
\]
and every $k_0$ spans an invariant eigenspace ${\cal H}_{k_0}$ of
dimension $(k_0 + 1)^2$.  The direct sum of these spaces leads to the
quasiregular representation of the isometry group $I_1 = SO(4)$ of
$\EuScript{S}_1$, i.e. by the theorem of Peter-Weyl the Hilbert space
of square integrable functions on $\EuScript{S}_1$ is given by
\[
   L^2(\EuScript{S}_1) = \bigoplus_{k_0=0}^\infty {\cal H}_{k_0} \, .
\]
The Fourier transform
\[
   \tilde{h}(\vec{k}) := (Y_{\vec{k}},h)
\]
gives a unitary transformation of $L^2(\EuScript{S}_1)$ to
$L^2(\tilde{\EuScript{S}}_1)$, where $\tilde{\EuScript{S}}_1$ denotes the
momentum space associated with $\EuScript{S}_1$, i.e. the range of values of
$\vec{k}$ equipped with the counting measure.

\subsection{The flat case}

The isometry group of the space $\EuScript{S}_0$ is the
Euclidean group $I_0 = E(3)$. The generalized eigenvectors
of the Laplace operator are
\[
   Y_{\vec{k}}(\vec{x}) = (2\pi)^{-3/2} e^{i\vec{k} \cdot \vec{x}} \, ,
\]
regarded as distributions over $\EuScript{S}_0$. They obey the equation
\[
   \Delta_0 Y_{\vec{k}}(\vec{x}) = -k^2 Y_{\vec{k}}(\vec{x}) \, ,
   \quad k := |\vec{k}| \, .
\]
We have the Fourier transform as a unitary mapping of $L^2(\EuScript{S}_0)$
to $L^2(\tilde{\EuScript{S}}_0)$ as usually given by
\[
    \tilde{h}(\vec{k}) := (Y_{\vec{k}},h) \, , \quad h \in
         L^2(\EuScript{S}_0)\, .
\]
With each $k \in {\mathbb R}_+$ we associate a function $h_k \in
C_0^\infty({\EuScript S}_0)$, by $k \to \tilde{h}_k$, taking values in
$L^2(S^2,d\Omega)$ ($S^2$ the two sphere and $d\Omega$ the invariant
measure on it):
\begin{equation}\label{fouriertrafo}
  \tilde{h}_k(\vec{\xi}\,) = \int d\mu(\vec{x})
       \overline{Y_{k\vec{\xi}}(\vec{x})} h(\vec{x}) \, .
\end{equation}
The map $h \to \tilde{h}$ extends to express the Hilbert space
$L^2({\EuScript S}_0)$ as a direct integral over ${\mathbb R}_+$:
\[
   L^2({\EuScript S}_0) = \int^{\oplus}  L^2(S^2,d\Omega) k^2 dk \, .
\]

\subsection{The hyperbolic case}

The isometry group $I_{-1}$ of the hyperbolic space ${\EuScript
S}_{-1}$ is the Lorentz group ${\mathcal L}_+^\uparrow(4)$. The
generalized eigenvectors of the Laplace operator $\Delta_{-1}$ on
this space are distributions of the form
\[
   Y_{\vec{k}}(\vec{x}) = (2\pi)^{-3/2} (x \cdot \xi)^{-1+ik} \, ,
          \quad \vec{k} = k\vec{\xi} \in {\mathbb R}^3,
\]
where $x \cdot \xi$ means the scalar product in Minkowski spacetime,
$\xi = (1,\vec{\xi}) \in {\mathbb R}^4$,
with eigenvalues given by
\[
   \Delta_{-1} Y_{\vec{k}}(\vec{x}) = -(k^2 + 1) Y_{\vec{k}}(\vec{x}) \, .
\]
Similar to the flat case we have a Fourier transformation given by
equation (\ref{fouriertrafo}) and a direct integral decomposition
of $L^2({\EuScript S}_{-1})$ (cf. \cite[Ch.VI 3.3]{Gelfandua} and
\cite[Ch.X \S 4]{Vilenkin}).

\subsection{The propagator}

As a consequence of the global hyperbolicity of the manifold $(\M,g)$
we have the existence of
global fundamental solutions of the Klein-Gordon equation
(see \cite{ChoquetBruhat}).
There exist unique operators $E^{\pm}:
C_0^\infty(\M) \to C^\infty(\M)$, satisfying
\begin{eqnarray*}
   (\Box_g - m^2)E^{\pm} &=& E^{\pm}(\Box_g - m^2) = I \, , \\
   \mbox{supp}(E^\pm \phi) &\subset& J^\pm (\mbox{supp} \phi) \, ,
   \quad \phi \in C_0^\infty(\M) \, ,
\end{eqnarray*}
where $J^+(p)$ is the causal future and $J^-(p)$ is the
causal past of $p \in \M$, i.e. all $q \in \M$ that can be joined
with $p$ by a future resp.~past directed causal curve and $I$
the identity. The operator $E := E^+ - E^-$ is the causal
propagator. This operator can be extended continuously to a mapping
from the space of distributions with compact support to the space
of distributions (on $\M$ resp.):
$E^{\pm}:{\cal E}'(\M) \to {\cal D}'(\M)$.

Since we know the eigenfunctions and eigenvalues of the operator
$\Delta_\varepsilon$ the propagator is determined by the ordinary
differential operator
\begin{equation}
        D = \frac{d^2}{dt^2} + 3 \frac{\dot{R}(t)}{R(t)} \,
         \frac{d}{dt} + m^2 + \frac{E(k)}{R^2(t)}  \, ,
\end{equation}
where $E(k)$ is a unified notation of the eigenvalues of the
Laplace operators $\Delta_\varepsilon$, i.e. $E(k)=k_0(k_0+2)$ for
$\varepsilon=1$, $E(k)=k^2$ for $\varepsilon=0$ and $E(k)=k^2+1$ for
$\varepsilon=-1$.

Unfortunately this differential equation can be solved
analytically only in very limited
cases explicitly, namely if the scaling factor $R(t)$ is
proportional to $t$. There are two fundamental solutions
in that case \cite[C.2.162]{Kamke}:
$J_\nu(mt)/t$ and $Y_\nu(mt)/t$, $\nu := \sqrt{1-E(k)}$, where
$J_\nu$ and $Y_\nu$ are Bessel functions of the first and second
kind respectively \cite{EMOT2}.

\section{The Weyl algebra}
\label{section_Weyl_algebra}

In this section we construct the Weyl algebra $CCR(D,\sigma)$
associated with the space of classical solutions of the Klein-Gordon
equation on Robertson-Walker spacetimes. To this end
we need a real symplectic vector space. There are two equivalent
methods for constructing an algebra on globally hyperbolic spacetimes
\cite{Dimock80}. One method is based
on the solutions on a Cauchy surface while the other one starts
with the vector space $\Cnull{\M}/\ker E$, where $\ker E$ is the
kernel of the propagator $E$. We use the first method, because it is
appropriate for our problem.

As the real symplectic vector space we take the space of real Cauchy
data on a Cauchy surface $D(\EuS) := \Cnull{\EuS} \oplus
\Cnull{\EuS}$. For a function $f \in \Cnull{\M}$ the restriction to
$D(\EuS)$ is given by $\rho_0 Ef \oplus \rho_1 Ef$, where $\rho_0:
\Cnull{\M} \to \Cnull{\EuS}$ is the restriction operator to the Cauchy
surface and $\rho_1: \Cnull{\M} \to \Cnull{\EuS}$ is the forward
normal derivative on the Cauchy surface. The Weyl algebra
$CCR(D,\sigma)$ is the algebra generated by the elements
$W(F) \neq 0, F \in D(\EuS)$,
obeying the Weyl form of the canonical commutation relations
\[
   W(F)W(G) = e^{-i\sigma(F,G)/2}W(F+G)\, , \quad F,G \in D(\EuS),
\]
and with the property $W(F)^* = W(-F)$, see e.g.~\cite[Ch.~8.2]
{BaumgartelWollenberg}. The symplectic form is given by
\[
   \sigma(F,G) = R^3 \int_{\EuS} (f_1g_2 - f_2g_1) \,d\mu(\EuS),
   \quad F=f_1\oplus f_2, \quad  G=g_1\oplus g_2 \, ,
\]
where the invariant measure on the respective spaces is meant.
The net of local observables is defined as
\[
   \A(\O) = C^*(W(\rho_0 Ef \oplus \rho_1 Ef), \; \mbox{supp}\;(\rho_i
        Ef) \subset \O \subset \M, \; i=0,1, \; f \in \Cnull{\M}) \; ,
\]
where we mean the $C^*$-algebra generated by these elements. We
define the quasilocal algebra by
\[
        \A = \ol{\bigcup_{\cal O}\A(\M)}  \, ,
\]
where $\O$ runs through all open subsets of \M.

This net fulfills a set of axioms formulated by Dimock
\cite{Dimock80}, generalizing the Haag-Kastler axioms (see
\cite{HaagKastler}) to curved
spacetimes. If we have a (regular) representation of $CCR(D,\sigma)$,
we can define a selfadjoint field operator $\varphi$ with the help of
Stone's theorem by
\[
        \exp[it\varphi(F)] = W(tF) \, , \quad F \in D, \quad t \in
        {\mathbb R}.
\]

\section{Adiabatic vacuum states}
\label{adiavacuum}

Adiabatic vacuum states were introduced by Parker
\cite{Parker69I} with the intention to determine the state in which
the creation rate of particles forced by expansion of the universe
is minimal. L\"uders and
Roberts analysed these states in the framework of algebraic quantum
field theory \cite{LuedersRoberts} and proved that the family of
adiabatic vacuum states is consistent with the principle of local
definiteness. This principle was introduced by Haag, Narnhofer and
Stein \cite{HaagNarn} in discussing quantum field theory on
curved spacetimes (cf.~also \cite[III.3.1]{Haag}) for to distinguish
the physically realizable states among all positive normalized
functionals which have no physical significance. Furthermore it is
now known
that adiabatic vacuum states are Hadamard states \cite{Junker}.

\subsection{Definition and construction of adiabatic vacua}

We recall the definition of quasifree and Fock states
(cf.~e.g.~\cite[8.2.3]{BaumgartelWollenberg}).  Let
$CCR(D,\sigma)$ be the Weyl algebra over the real vector space $D$
with symplectic form $\sigma$ and let $S(\cdot,\cdot)$ be a real scalar
product on $D$, satisfying
\[
   |\sigma(F,G)| \le \sqrt{S(F,F)}\sqrt{S(G,G)} \, , \quad F,G \in D.
\]
A quasifree state (with vanishing one-point function) is defined
by the generating functional
\[
   \omega_S(W(F)) = \exp(-S(F,F)/4)\, , \quad F \in D,
\]
where $W(F)$ denotes the Weyl operator. Such a quasifree state
$\omega_S$ determines the two-point function
\[
   \skalar{\Omega}{\varphi(F)\varphi(G)\Omega}_S = [S(F,G) + i
   \sigma(F,G)]/2 \, , \quad F,G \in D,
\]
where $\varphi(F)$ is the generator of the Weyl operator in the
GNS-representation constructed of $\omega_S$ and $\Omega$
the corresponding GNS-vacuum.

If the scalar product $S$ can be linked with the symplectic form
$\sigma$ by an internal complexification $J$, i.e. $J^2 = -1$,
$\sigma(JF,G) = -\sigma(F,JG)$, $\sigma(F,JG) = S(F,G)$, the
state is called a Fock state.

We now recall the construction, definition and some properties of
adiabatic vacuum states from L\"uders and Roberts \cite{LuedersRoberts}.
In their analysis of the structure of states on a \RW they defined a
quasifree state $\omega$ to be homogeneous and isotropic, if $\omega
\circ \alpha_g = \omega$, where $g$ is an element of the isometry
group $I_\varepsilon$ of the spaces $\EuS$ and $\alpha_g$ the
corresponding automorphism. The analysis of these conditions
leads to the function spaces described in
section \ref{section_classical}. For to find conditions on the scalar
product $\skalar{\cdot}{\cdot}_S$ to define a Fock state
they imposed the following continuity condition.
There is a $\nu \in {\mathbb N}_0$ and a constant $C > 0$, such that
\begin{equation}\label{stetbed}
  \skalar{F}{G}_S \le C\|F\|_\nu\|G\|_\nu\, , \quad F,G \in D,
\end{equation}
where
\begin{eqnarray*}
  \|F\|^2_\nu &:=& (F,(m^2-\Delta_{\varepsilon})^{2\nu}F) \, , \quad
  \nu \in \mathbb{N}_0 \, ,                                   \\
  (F,G) &:=& R^3\int_{\EuS} (f_1g_1 + f_2g_2) \, d\mu(\EuS) \, ,
  \quad F = f_1 \oplus f_2, \quad G = g_1\oplus g_2 \, .
\end{eqnarray*}
We denote the classical solutions
with the uniform notation
\begin{eqnarray}\label{f1f2}
  F &=& \binom{f_1}{f_2} =
  \int d\vec{k} \binom{c(\vec{k})}{\hat{c}(\vec{k})}
  Y_{\vec{k}}(\vec{x}) \, , \\ G &=& \binom{g_1}{g_2} = \int d\vec{k}
  \binom{b(\vec{k})}{\hat{b}(\vec{k})} Y_{\vec{k}}(\vec{x}) \, ,
  \quad F,G \in D \, .
  \nonumber
\end{eqnarray}
where the integral reduces to a sum in the spherical case. The main
result concerning the structure of Fock states is summarized in the
following
\begin{theo}
  The homogeneous and isotropic Fock states for the Klein-Gordon field
  on a \RW fulfilling the continuity condition (\ref{stetbed}) are
  given by a two-point function of the form
  \begin{eqnarray}\label{zpfunk}
    \skalar{F}{G}_S &=& \int d\vec{k} [ c(\vec{k})b(\vec{k}) S_{00}(k) +
    c(\vec{k})\hat{b}(\vec{k}) S_{01}(k) \nonumber \\ & & {}+
    b(\vec{k})\hat{c}(\vec{k}) S_{10}(k) +
    \hat{c}(\vec{k})\hat{b}(\vec{k}) S_{11}(k) ] \, .
  \end{eqnarray}
  The entries of the matrix $S$ can be expressed in the form
  \begin{eqnarray}\label{matrix}
    S_{00}(k) &=& |q(k)|^2 \, , \qquad
    S_{11}(k) = |p(k)|^2 \, , \nonumber \\ S_{01}(k) &=& \ol{q(k)}
    p(k) \, , \qquad S_{10} = \ol{S}_{01} \, ,
  \end{eqnarray}
  where $p$ and $q$ are essentially polynomially bounded measurable
  functions satisfying
  \begin{equation}\label{bedi}
    \ol{q(k)}p(k) - \ol{p(k)}q(k) = i \, .
  \end{equation}
  Conversely every pair of polynomially bounded measurable functions
  satisfying equation (\ref{bedi}) yields via (\ref{matrix}) and
  (\ref{zpfunk}) the two-point function of a homogeneous, isotropic
  Fock state which satisfies the continuity condition (\ref{stetbed}).
\end{theo}
For the proof see \cite[Thm.~2.3]{LuedersRoberts}.

The dynamical evolution of the states is expressed by the
coefficients $c(\vec{k})$ and $\hat{c}(\vec{k})$ in equation
(\ref{f1f2}). An adiabatic vacuum state is
given by a specific initial condition on a Cauchy surface which
fixes the matrix $S$. This means that the adiabatic vacuum state
depends on the Cauchy surface.

The dynamical equation is
\begin{equation}\label{t2kg}
   \left( \frac{d^2}{dt^2} + 3 \frac{\dot{R}(t)}{R(t)} \,
   \frac{d}{dt} + m^2 + \frac{E(k)}{R^2(t)} \right)
    c_k(t) = 0 , \quad \forall k \, .
\end{equation}
This equation can be solved explicitly only in exceptional cases. In
the general case, one tries to solve it by an iteration procedure.
For to find the iteration, we consider
\[
        T_{k}(t) = [2R^3(t)\Omega_k(t)]^{-1/2} \exp
                \left(i \int_{t_0}^t \Omega_k(t') \, dt' \right) \, ,
                \quad \forall k \, ,
\]
where the functions $\Omega_k$ have to be determined. Inserting this
ansatz in equation (\ref{t2kg}) we find that the functions $\Omega_k$
have to satisfy
\[
        \Omega_k^2 = \omega_k^2 - \frac{3}{4}
                \left( \frac{\ddot{R}}{R} \right)^2 - \frac{3}{2}
                \frac{{\dot R}}{R} + \frac{3}{4}
                \left( \frac{\dot{\Omega}_k}{\Omega_k} \right)^2
                -\frac{1}{2} \frac{\ddot{\Omega}_k}{\Omega_k} \, ,
\]
where $\omega_k^2 = E(k)/R^2 + m^2$. With
\[
    (\Omega_k^{(0)})^2 := \omega_k^2 = E(k)/R^2 + m^2
\]
the iteration is given by
\[
        (\Omega_k^{(n+1)})^2 = \omega_k^2 - \frac{3}{4}
                \left( \frac{\ddot{R}}{R} \right)^2 - \frac{3}{2}
                \frac{{\dot R}}{R} + \frac{3}{4}
                \left( \frac{\dot{\Omega}_k^{(n)}}{\Omega_k^{(n)}}
                \right)^2 -\frac{1}{2} \frac{\ddot{\Omega}_k^{(n)}}
                {\Omega_k^{(n)}}     \, .
\]
The functions $T_k(t)$ and $\dot{T}_k(t)$ are related to the
functions $q(k)$ and $p(k)$, which constitute the matrix $S$
by equation (\ref{matrix}). On a Cauchy surface at time $t$
these relations are
\begin{equation}\label{adia1}
  T_k(t) = q(k) \, , \quad \dot{T}_k(t) = R^{-3}(t)p(k)  \, .
\end{equation}
An adiabatic vacuum state will now be defined by initial values
at time $t$:
\begin{defi}
  For $t_0,t \in {\mathbb R}$, let
  \begin{equation}\label{adiabat}
    W_k^{(n)}(t) := [2R^3(t)\Omega_k^{(n)}(t)]^{-1/2}
    \exp \left(i \int_{t_0}^t \Omega_k^{(n)}(t') \, dt' \right) \, .
  \end{equation}
  An adiabatic vacuum state of order $n$ is a Fock state, obtained via
  equations (\ref{matrix}) and (\ref{adia1}), where the initial values
  at time $t$ for equation (\ref{t2kg}) can be expressed by
  \[
    T_{k}(t) =      W_k^{(n)}(t) \, , \qquad
    \dot{T}_{k}(t) = \dot{W}_k^{(n)}(t) \, .
  \]
\end{defi}
These states depend on the initial time $t$ in equation
(\ref{adiabat}), the order of iteration $n$ and the extrapolation of
$\Omega^{(n)}_k$ to small values of $k$. The main result on the
family of adiabatic vacuum states is summarized (see L\"uders and
Roberts \cite[Thm.~3.3]{LuedersRoberts}) in
\begin{theo}
  In a closed \RW any two adiabatic vacuum states for the free
  Klein-Gordon field are unitarily equivalent.\\
  In a flat or hyperbolic \RW any two adiabatic vacuum states of order
  $n \ge 1$ for the free Klein-Gordon field are locally
  quasiequivalent.
\end{theo}
A generalization to $n \ge 0$ also in the flat and hyperbolic case was
given by Junker \cite[Cor.~3.23]{Junker}, so that  all adiabatic
vacuum states are locally quasiequivalent and we can perform our
computations in an adiabatic vacuum state of order zero.

Local quasiequivalence of two states $\omega$ and $\omega'$ means that
the density matrices in the GNS Hilbert spaces ${\cal H}_\omega$ and
${\cal H}_{\omega'}$ define the same set of states of $\A(\O)$ for
each $\O$ in $\M$. The principle of local definiteness mentioned
in section \ref{section_Weyl_algebra}
requires local quasiequivalence of the physical realizable states.

The entries of the matrix $S$ for a zeroth order adiabatic vacuum
state are given in terms of $\Omega \equiv \Omega_k^{(0)}$ by
\begin{eqnarray*}
  S_{00}(k) &=& |q(k)|^2 = \frac{1}{2\Omega R^3} \\
  S_{01}(k) &=& \overline{q(k)}p(k) = \frac{i}{2} -
  \frac{3\dot{R}}{4\Omega R} - \frac{\dot{\Omega}}{4\Omega^2} \\
  S_{11}(k) &=& |p(k)|^2 = \frac{\Omega R^3}{2} +
  \frac{9R\dot{R}^2}{8\Omega} + \frac{3\dot{\Omega}R^2\dot{R}} {4\Omega^2}
  + \frac{R^3\dot{\Omega}^2}{8\Omega^3} \, .
\end{eqnarray*}
The internal complexification $J$ on $D$ which leads to adiabatic vacuum
states is given by
\[
   J = \begin{pmatrix}
          \frac{3}{2}\frac{\dot{R}}{\Omega R} +
          \frac{\dot{\Omega}}{2\Omega^2} & -\Omega R^3 - \frac{9}{4}
          \frac{R \dot{R}^2}{\Omega} - \frac{3}{2}
          \frac{\dot{\Omega}R^2 \dot{R}}{\Omega^2} -
          \frac{R^2 \dot{\Omega}^2}{4\Omega^3}                      \\
          \frac{1}{R^3 \Omega} & - \frac{3}{2}
          \frac{\dot{R}}{\Omega R} - \frac{\dot{\Omega}}{2 \Omega^2}
       \end{pmatrix} \, .
\]

\subsection{The field operator}

We can define a field operator on the symmetrized
Fock space ${\cal F}(L^2(\EuS))$
in the usual way as an operator valued
distribution following Dimock \cite{Dimock80} and L\"uders and
Roberts \cite{LuedersRoberts}. In our case the field operator is
\[
   \varphi(f) = \theta(\rho_1 Ef) - \pi(\rho_0 Ef) \, , \quad f
   \in \Cnull{\M} \, .
\]
with time-zero fields
\begin{eqnarray*}\label{fieldoperator}
  \theta(g) &=& a(qR^3 \ol{\tilde{g}}) + a^*(qR^3 \tilde{g}) \,
  , \\ \pi(g) &=& a(p \ol{\tilde{g}}) + a^*(p \tilde{g})
  \, , \quad g \in \Cnull{\EuS} \, ,
\end{eqnarray*}
where $a^*$ and $a$ are the usual creation and annihilation operators.
Hence
\[
     \varphi(f) = a(qR^3 \, \ol{(\rho_1 Ef)\tilde{}} \, )
           + a^*(qR^3 \, (\rho_1Ef)\tilde{ } \, )
           - a(p \, \ol{(\rho_0 Ef)\tilde{}} \,)
           - a^*(p \, (\rho_0 Ef)\tilde{ } \,) \, .
\]

\section{Gaussian measures and the space $L^2(\D'(\N))$}

The spectral decomposition of the field operator can be achieved on a
functional Hilbert space over the distributions ${\cal D}^\prime(\N)$,
$\N$ being a manifold. The inner product on this space is determined
by choice of a Gaussian measure $\mu_C$ with
covariance $C$. We will
outline the construction of the measure (cf.~e.g.~Gelfand and
Wilenkin \cite[Ch.~IV]{GelfandWilenkin} for details).

Let $X$ be a locally convex topological vector space and $C$ a scalar
product on $X$, the covariance. For every finite-dimensional subspace
$U$ of $X$ with
$\dim U = n$ the Gaussian measure $\sigma_U$ is defined by
\[
     \sigma_U(A) := (2\pi)^{-n/2} \int_A \exp[-C(x,x)/2] dx \, ,
     \quad A \subset U,
\]
where $dx$ is the Lebesgue measure on $U$ with respect
to the scalar product $C$. With help of the natural isomorphism
$P:U \to X'/U^\perp$, where $U^\perp = \{ F \in X' | F(x_i) = 0
\;(i=1,\ldots,n), x_i \; \mbox{span} \; U\}$ and $X'$ the topological
dual of $X$, we define a family of
measures $\nu_U$ on $X'/U^\perp$ by
\[
   \nu_U(A) = \sigma_U(P^{-1}(A))\, , \quad A \subset X'/U^\perp.
\]
This family of measures $\nu_U$ induces a measure $\mu$ on the
cylinder sets $\cal Z$ of $X'$ by
\[
   \mu(Z(U,A)) = \nu_U(A) \, ,
\]
where $Z(U,A) \in \cal{Z}$ is the cylinder set with base $A$ and
generating space $U^\perp$, i.e. $Z(U,A) = \{F \in X'|
(F(x_1),\ldots,F(x_n)) \in A,\; x_i \; \mbox{span} \; U \; ( i
=1,\ldots,n)\}$.
The measure can be extended to a measure on the
$\sigma$-algebra generated by the cylinder sets if it is
$\sigma$-additive. The following theorem states that $X$ has to be
nuclear.
\begin{theo}
  Sufficient and necessary for a measure on the dual space $X'$ of a
  locally convex vector space $X$ to be $\sigma$-additive is the
  nuclearity of the space $X$.
\end{theo}
For the proof see Gelfand and Wilenkin \cite[Ch.~IV, \S
3]{GelfandWilenkin}.

Hence, the construction presupposes that $\D(\N)$ is nuclear. A
sufficient condition is given by the next theorem (see Maurin
\cite[I.9]{Maurin2}).
\begin{theo}
The space $\D(\N)$ is nuclear, if
$\N$ is a $\sigma$-compact manifold.
\end{theo}
This means $\N$ is the countable union of compact subspaces.
This condition is not too restrictive.
For example every separable and locally compact manifold
is $\sigma$-compact and especially the spaces $\D(\EuS)$ are nuclear.

\section{Spectral resolution and correlations}

For the spectral decomposition of the field operator we choose the
functional Hilbert space $L^2(\D'(\EuS), \Sigma,\mu_{1/2})$, where
$\D'(\EuS)$ is the space of distributions over $\EuS$, $\mu_{1/2}$ the
Gaussian measure on the $\sigma$-algebra $\Sigma$ generated by the
cylinder sets of $\D'(\EuS)$ with covariance $C(f,g) =
\skalar{f}{g}/2$, $\skalar{f}{g}$ is the scalar product of $L^2(\EuS)$. A
complete orthonormal system on $L^2(\D'(\EuS))$ is given in terms of
the Hermite polynomials $H_n$ by $2^{-n/2}n!\,H_n(2^{1/2}\Phi(f)), n
\in {\mathbb N}_0, f \in \D(\EuS), \|f\| = 1$
(see \cite[Ch.~6.3]{GlimmJaffe}).
The time-zero field $\theta(f)$ (see equation
(\ref{fieldoperator})) acting on the Fock space ${\cal F}(L^2(\EuS))$
can be represented as an operator of
multiplication by $\sqrt{2}\Phi(f)$ on the Hilbert space $L^2({\cal
  D}'(\EuS))$. The unitary transformation $U: {\cal F}(L^2(\EuS)) \to
L^2({\cal D}'(\EuS))$, is given by
\[
   U: \underbrace{f \otimes \ldots \otimes f}_{n-times} \mapsto
    (n!)^{-1/2} H_n(2^{1/2}\Phi(f)) \, , \quad \|f\| = 1.
\]

Having represented the field operator $\theta$ as a multiplication
operator on $L^2(\D'(\EuS),\mu_{1/2})$, we are able to compute the
probability distribution for a field operator $\phi(f)$ or
correlations for a pair of field operators $\phi(f), \phi(g)$ as a
consequence of a certain field configuration.

We compute these correlations in the time-zero field $\theta$
because it is not possible to represent $\theta$ and $\pi$ on
$L^2(\D'(\EuS),\mu_{1/2})$ simultaneously as multiplication operators.
But if we find correlations in the time-zero field $\theta$, they
will also be present in the field operator $\varphi = \theta - \pi$.

The spectral projections $P_\Delta(\varphi)$ of an operator $\varphi$
($\Delta \subset {\mathbb R}$ a Borel set) are given by
$P_\Delta(\varphi) \equiv \chi_\Delta(\varphi)$, where $\chi_\Delta$
is the characteristic function of the set $\Delta$.  Denoting by
$M_{\chi_\Delta(\Phi)}$ the operator on $L^2(\D'(\EuS),\mu_{1/2})$
which is defined by multiplication with the function
$\chi_\Delta(\Phi)$,  we have $\chi_\Delta(\theta(f)) =
U^{-1}M_{\chi_\Delta(\sqrt{2}\Phi(f))}U$ and with $U\Omega = 1$
\begin{eqnarray*}
   \skalar{\Omega}{\chi_\Delta(\theta(f))\Omega} &=&
      \skalar{\Omega}{U^{-1}M_{\chi_\Delta(\sqrt{2}\Phi(f))}U\Omega} \\
      &=& \sqrt{2}\int_\Delta\Phi(f) d\mu_{1/2} \, .
\end{eqnarray*}
These Gaussian integrals can be solved explicitly.
The probability measures $\mu_C$ are given by
(see Gelfand and Wilenkin \cite[Ch.\ 3, \S 3]{GelfandWilenkin})
\begin{multline*}
        \mu_C\{ \Phi(f_1)=x_1, \ldots , \Phi(f_n)=x_n \} = \\
               = (2\pi)^{-n/2} (\det C)^{-1/2} \int \exp(
                -\frac{1}{2}\sum_{i,j=1}^n (C^{-1})_{ij}x_ix_j)
                \prod_{i=1}^n  \, dx_i \, ,
\end{multline*}
where $C=((C_{ij}))$, $(i,j=1,\ldots,n)$ is the covariance matrix.
In our case the entries are
\[
        C_{ij} := \skalar{f_i}{f_j}/2 \,, \quad f_i
        \in {\cal D}(\EuS), \quad  i,j=1,\ldots,n \, ,
\]
such that
\[
        \int d\mu_{1/2} = (2\pi c)^{-1/2} \int \exp (-c^2x^2/2)dx \, ,
                \quad c^2:=2/\|f\|^2 \, ,
\]
and for an interval $\Delta = [a,b] \in {\mathbb R}$,
$f \in {\cal D}(\EuS)$, we have
\begin{eqnarray}\label{korr3}
  \int_a^b \Phi(f) \, d\mu_{1/2}
     &=&  c(2\pi)^{-1/2} \int_a^b x \exp(-c^2x^2/2) \, dx   \nonumber \\
     &=&  \frac{\|f\|}{2 \sqrt{\pi}} [ \exp(-a^2\|f\|^{-2})
       - \exp(-b^2\|f\|^{-2}) ]\, .
\end{eqnarray}
For the product $\Phi(f)^2, f \in {\cal D}(\EuS)$, we find
\begin{eqnarray*}
   \int_a^b \Phi(f)^2 \, d\mu_{1/2}
      &=&  \frac{\|f\|}{2\sqrt{\pi}}  [ a \exp(-a^2\|f\|^{-2})
           - b \exp(-b^2\|f\|^{-2}) ]       \\
      & &  {}+ \frac{\|f\|^2}{4} [ \mbox{Erf}\:(b\|f\|^{-1})
           - \mbox{Erf}\:(a\|f\|^{-1})]          \, .
\end{eqnarray*}
where we have defined the integral
\[
   \mbox{Erf}\:(d) := \frac{2}{\sqrt{\pi}} \int_0^d \exp(-x^2) \, dx
                \, , \quad d \in {\mathbb R}.
\]
With the help of Wicks theorem we find that $c =
2\skalar{f}{g}^{-1/2}$ and for the product of two field operators
the formulae are just
\begin{eqnarray}\label{korr4}
   \int_a^b \Phi(f) \Phi(g) \, d\mu_{1/2}
      &=&       \frac{\skalar{f}{g}^{1/2}}{2\sqrt{\pi}} [
         a \exp ( -a^2\skalar{f}{g}^{-1}   ) -
         b \exp ( -b^2\skalar{f}{g}^{-1}   )]       \nonumber \\
      & & {}+ \frac{\skalar{f}{g}}{4}
   [ \mbox{Erf}\: ( b\skalar{f}{g}^{-1/2}) -
     \mbox{Erf}\:( a\skalar{f}{g}^{-1/2})] \, .
\end{eqnarray}
We are now ready to compute the expectation value
\[
   <\chi_\Delta(\theta(f) \theta(g))> \;\;
     = 2 \int_\Delta \Phi(\,qR^3 \, (\rho_1 Ef)\tilde{} \, )
        \Phi(\, qR^3\, (\rho_1 Eg)\tilde{} \, ) \, d\mu_{1/2} \, ,
\]
where we can use formula (\ref{korr4}) now. On the other hand we have
\[
   <\chi_\Delta(\theta(f))><\chi_\Delta(\theta(g))> \;\;
     = 2\int_\Delta \Phi(\,qR^3 \, (\rho_1 Ef)\tilde{} \, )
        \int_\Delta \Phi(\,qR^3 \, (\rho_1 Eg)\tilde{} \, ) \, ,
\]
where we can use formula (\ref{korr3}) now.

We summarize some formulae:
The expectation value of the product of two field operators
to have a value in the interval $\Delta = [a,b]$ is given by
\begin{eqnarray*}
  \lefteqn{ <\chi_\Delta(\theta(f) \theta(g))> \;\;
     = 2 \int_\Delta \Phi(\,qR^3 \, (\rho_1 Ef)\tilde{} \, )
        \Phi(\, qR^3 \, (\rho_1 Eg)\tilde{} \, ) \, d\mu_{1/2}(\Phi)} \\
     &=& \skalar{qR^3 \,(\rho_1 Ef)\tilde{}\,}
                  {qR^3 \,(\rho_1 Eg)\tilde{}\,}^{1/2}\pi^{-1/2}
         [ a \exp ( -a^2/\skalar{qR^3 \,(\rho_1 Ef)\tilde{}\,}
                              {qR^3 \,(\rho_1 Eg)\tilde{}\,})         \\
     & & {}- b \exp ( -b^2/\skalar{qR^3 \,(\rho_1 Ef)\tilde{}\,}
                              {qR^3 \,(\rho_1 Eg)\tilde{}\,}   )]      \\
     & & {}+ 2^{-1}\skalar{q R^3 \, (\rho_1 Ef)\tilde{} \,}
                   {q R^3 \,(\rho_1 Eg)\tilde{}\,}
         [\mbox{Erf}\: ( b\skalar{qR^3 \,(\rho_1 Ef)\tilde{}\,}
                            {qR^3 \,(\rho_1 Eg)\tilde{}\,}^{-1/2}) \\
     & & {}- \mbox{Erf}\:( a\skalar{qR^3 \,(\rho_1 Ef)\tilde{}\,}
                           {qR^3 \,(\rho_1 Eg)\tilde{}\,}^{-1/2})] \, .
\end{eqnarray*}
On the other hand we have
\begin{eqnarray*}
   \lefteqn{<\chi_\Delta(\theta(f))><\chi_\Delta(\theta(g))> =} \\
    &=& 2\int_\Delta \Phi(\,qR^3 \, (\rho_1 Ef)\tilde{} \, )
        \int_\Delta \Phi(\,qR^3 \, (\rho_1 Eg)\tilde{} \, ) =   \\
    &=&  2\pi^{-1}\|qR^3 \,(\rho_1 Ef)\tilde{}\,\|
          \|qR^3 \,(\rho_1 Eg)\tilde{}\,\|
          [ \exp(-a^2\|(\rho_1 Ef)\tilde{}\,\|^{-2})
       - \exp(-b^2\|qR^3 \,(\rho_1 Ef)\tilde{}\,\|^{-2}) ]             \\
    & & {}   [ \exp(-a^2\|qR^3 \,(\rho_1 Eg)\tilde{}\,\|^{-2})
       - \exp(-b^2\|qR^3 \,(\rho_1 Ef)\tilde{}\,\|^{-2}) ]\, .
\end{eqnarray*}
Easier formulae are obtained by taking
the interval $\Delta =[0,\infty )$
\begin{eqnarray}\label{formeln}
   <\chi_\Delta(\theta(f))><\chi_\Delta(\theta(g))> \;\; &=&
    \frac{1}{2\pi}\|qR^3 \,(\rho_1 Ef)\tilde{}\,\| \|qR^3 \,
    (\rho_1 Eg)\tilde{}\,\|               \nonumber              \\
    <\chi_\Delta(\theta(f) \theta(g))> \;\; &=&
    \frac{1}{2} \skalar{qR^3 \,(\rho_1 Ef)\tilde{}\,}
                  {qR^3 \,(\rho_1 Eg)\tilde{}\,} \, .
\end{eqnarray}

\noindent
{\bf Remark:} The Hilbert space is built over ${\cal D}'(\EuS)$, hence
one would expect that $\Cnull{\EuS}$-functions are needed in explicit
calculations. But since we are integrating
only over cylindrical sets and the propagator can be extended to the
spaces of distributions, it is possible to use $L^2(\EuS)$-functions
as well.  Nevertheless, by the complexity of the formulae, it is
impossible to evaluate them analytically, even in the case when
the propagator $E$ is explicitly known. We will analyse the formulae
(\ref{formeln}) in a forthcoming paper.\\[0.5cm]
{\bf Acknowledgments.} A major part of this work is taken from the
authors diploma thesis written at the Technische Universit\"at Berlin.
I want to thank Prof.~Hellwig for supervision and advice and M.~Keyl
for many helpful hints and discussions.


\begin{thebibliography}{99}
\bibitem{Wald2} R.~M.~Wald, Correlations Beyond the Horizon,
        Gen.~Rel.~Grav.~{\bf 24 }, 1111, (1992)
\bibitem{EMOT2} Erd\'{e}lyi et al., Higher Transcendental Functions
        Vol.~2, McGraw-Hill (1953)
\bibitem{Gelfandua} I.~M.~Gelfand, M.~I.~Graev and N.~Ya.~Vilenkin,
        Generalized Functions V, Academic Press, New York and London
        (1966)
\bibitem{Vilenkin} N.~J.~Vilenkin, Special Functions and the Theory of
        Group Representations, American Mathematical Society (1968)
\bibitem{ChoquetBruhat} Y.~Choquet-Bruhat, Hyperbolic Partial Diffential
        Operators, in: Battelle Rencontres, eds. deWitt and Wheeler,
        Benjamin (1968)
\bibitem{Kamke} E.~Kamke, Differentialgleichungen, Akademische
        Verlagsgesellschaften (1942)
\bibitem{Dimock80} J.~Dimock, Algebras of Local Observables on a
        Manifold, Commun.~Math.~Phys.~{\bf 77}, 219--228, (1980)
\bibitem{BaumgartelWollenberg} H.~Baumg\"artel and M.~Wollenberg,
        Causal Nets of Operator Algebras, Akademie Verlag (1992)
\bibitem{HaagKastler} R.~Haag and D.~Kastler,
        An Algebraic Approach to Quantum Field Theory,
        J.~Math.~Phys. {\bf 5}, 848--861, 1964
\bibitem{Parker69I} L.~Parker, Quantized Fields and Particle
        Creation in Expanding Universes {I},
        Phys. Rev. {\bf 183}, 1057--1068, (1969)
\bibitem{LuedersRoberts} C.~L\"uders and J.~E.~Roberts,
        Local Quasiequivalence and Adiabatic Vacuum States,
        Commun.~Math.~Phys.~{\bf 134}, 29--63, (1990)
\bibitem{HaagNarn} R.~Haag, H.~Narnhofer and U.~Stein,
        On Quantum Field Theory in Gravitational Background,
        Commun.~Math.~Phys.~{\bf 94}, 219--238, (1984)
\bibitem{Haag} R.~Haag, Local Quantum Physics, Springer-Verlag (1992)
\bibitem{Junker} W.~Junker, Adiabatic Vacua and Hadamard States for
        Scalar Quantum Fields in Curved Spacetime, PhD Thesis,
        Universit\"at Hamburg, (1995), DESY-preprint 95-144, hep-th
        9507097
\bibitem{GelfandWilenkin} I.~M.~Gelfand und N.~J.~Wilenkin,
        Verallgemeinerte Funktionen IV, VEB Deutscher Verlag der
        Wissenschaften (1964)
\bibitem{Maurin2} K.~Maurin, General Eigenfunction Expansions
        and Unitary Representations of Topological Groups,
        Polish Scientific Publishers (1968)
\bibitem{GlimmJaffe} J.~Glimm and A.~Jaffe,
        Quantum Physics. A Functional Integral Point of View,
        Springer Verlag (1987)
\end{thebibliography}
\end{document}